\documentclass[11pt,twoside]{article}

\usepackage{asp2004}
\usepackage{epsf}
\usepackage{psfig}
\usepackage{lscape}

\begin{document}
\title{The Quest for a Complete Census of AGN Activity}   
\author{D.M. Alexander}   
\affil{Department of Physics, Durham University, Durham DH1 3LE, UK}    

\begin{abstract} 
Deep X-ray surveys provide the most efficient identification of Active
Galactic Nuclei (AGN) activity. However, recent evidence has indicated
that the current $<10$~keV surveys do not detect the most heavily
obscured AGNs. Here we explore whether the X-ray undetected AGN
population can be identified within the ultra-deep {\it Spitzer}
survey of the GOODS-N field using X-ray stacking techniques. We find
evidence for AGN activity in the {\it Spitzer} dataset and the
strongest and hardest X-ray signal is produced by galaxies with
starburst-like infrared spectral slopes and median properties of
$z\approx$~0.8 and $L_{\rm IR}\approx10^{11}$~$L_{\odot}$. The
stacked X-ray properties suggest that obscured AGN activity is present
in these sources, with a median X-ray spectral slope of
$\Gamma\approx1$ and $L_{\rm X}\approx10^{42}$~erg~s$^{-1}$. These
overall properties are consistent with the obscured AGN population
expected to produce the unresolved X-ray background.

\end{abstract}


\section{Introduction}

There is a growing need for a complete census of AGN activity. The
seminal discovery that every massive galaxy in the local Universe
harbors a super-massive black hole (SMBH; $M_{\rm
BH}>10^{6}$~$M_{\odot}$) implies that all galaxies have hosted AGN
activity at some time during the past $\approx$~13~Gyrs (e.g.,\
\citealt{rees84}; \citealt{korm95}). To accurately trace how and when
these SMBHs grew requires a detailed census of AGN activity that will
provide, amongst other things, constraints on the efficiency and duty
cycle of SMBH growth. The finding that the mass of the SMBH is
proportional to that of the galaxy spheroid also indicates that the
growth of SMBHs and their host galaxies are regulated in some way
(e.g.,\ \citealt{mag98}; \citealt{gebh00}). Whether or not this growth
occured concordantly is unknown but comparisons between the rate of
star formation and SMBH growth, both in individual objects and from a
cosmic census, will provide important insight.

\begin{figure}[!t]
\plottwo{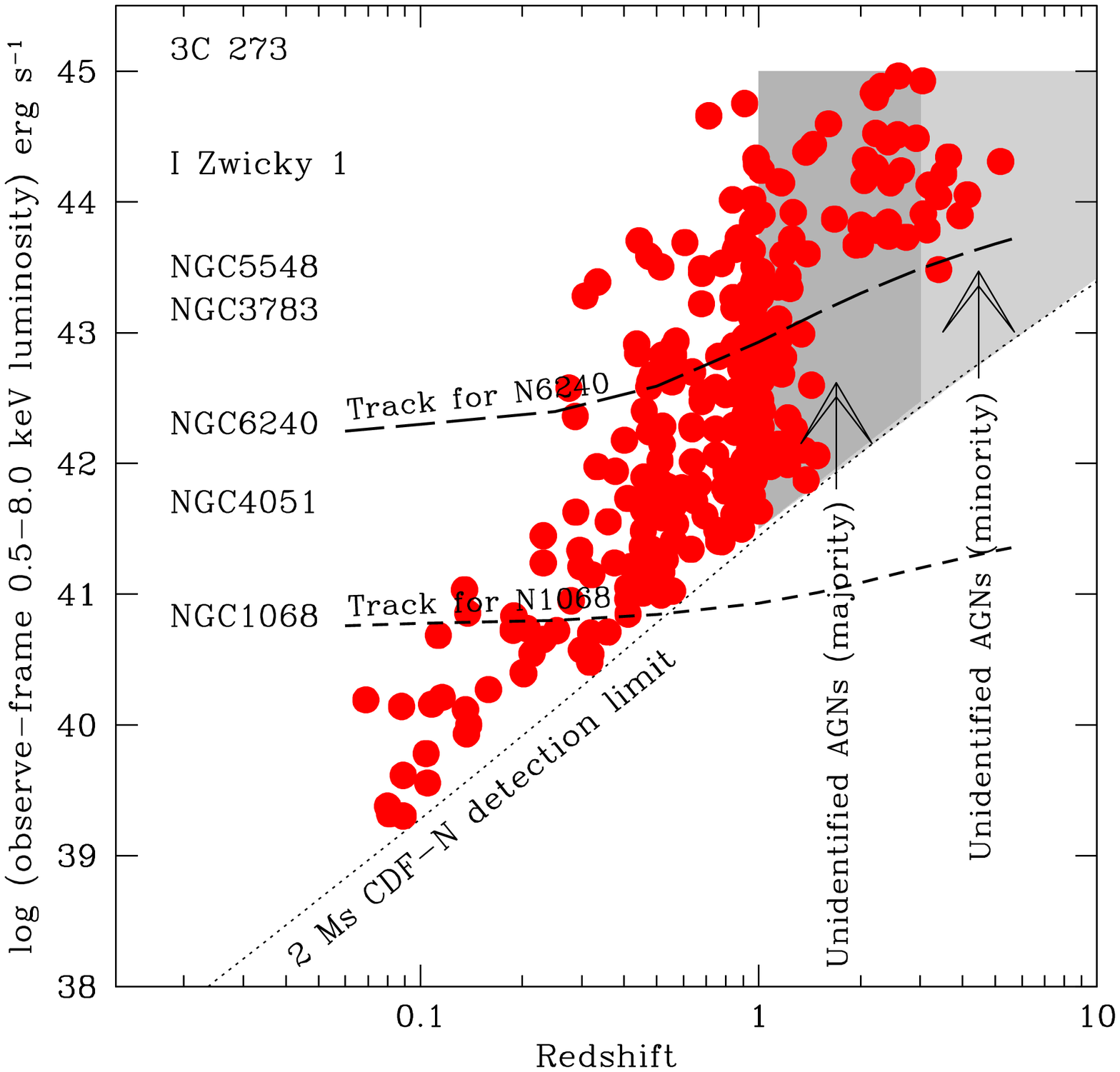}{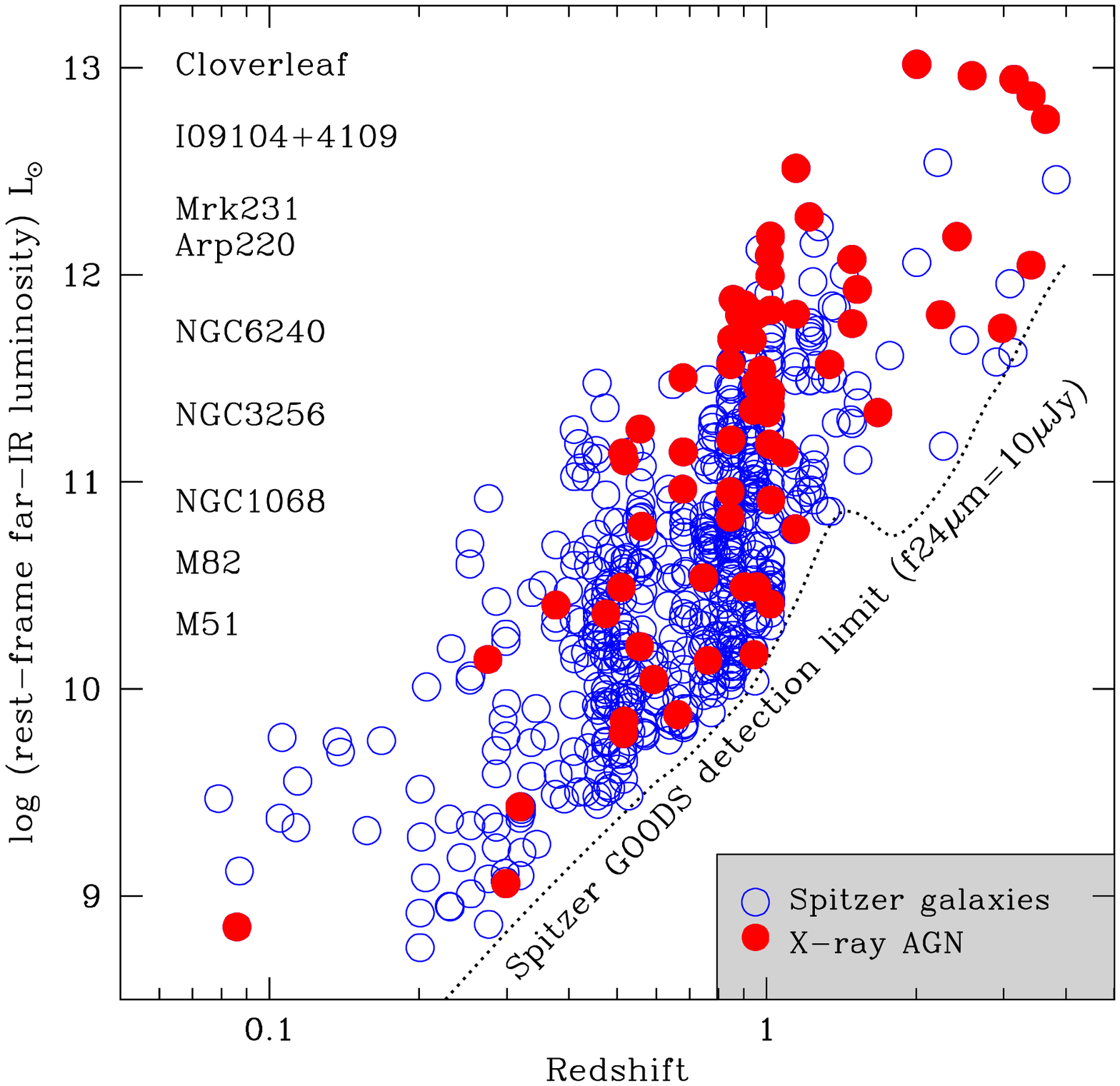}
\caption{(a) observed X-ray luminosity vs redshift for the
  spectroscopically identified X-ray sources in the CDF-N/GOODS-N
  field.  The tracks of the Compton-thick but intrinsically luminous
  ($L_{\rm X}\approx10^{44}$~erg~s$^{-1}$) AGNs in NGC~6240 (long
  dash) and NGC~1068 (short dash) are shown to indicate their
  detectability at different redshifts. The shaded regions show where
  the spectroscopically unidentified X-ray sources are likely to lie;
  $\approx$~50\% of the X-ray AGNs (almost all at $z>1$; e.g.,\
  \citealt{alex01}; \citealt{main05}) still lack spectroscopic
  redshifts. (b) rest-frame far-IR luminosity vs redshift for
  spectroscopically identified 24~$\mu$m {\it Spitzer} sources in the
  GOODS-N field; the far-IR luminosities are calculated from the
  24~$\mu$m data using the average of four SED templates (M~51, M~82,
  Arp~220, Mkn~231). Note that while NGC~1068 will only be detected
  out to $z\approx$~0.4 in the X-ray observations, it should be
  detectable out to $z\approx$~1--2 in the {\it Spitzer}
  observations.}
\end{figure}

\section{Selecting and Identifying AGN Activity}

Since obscured AGNs outnumber unobscured AGNs by a factor of
$\approx$~3--10, any ``complete'' census of AGN activity must be
sensitive to the detection and identification of obscured
sources. X-ray emission is relatively impervious to obscuration, at
least at high energies, and the detection of luminous or hard X-ray
emission provides an almost unambiguous identification of AGN activity
even in the absence of source redshifts. For this reason, the most
complete census of AGN activity to date is provided by deep X-ray
surveys. For example, the 2~Ms {\it Chandra} Deep Field North (CDF-N;
\citealt{alex03}) survey identifies a $>10$ times larger AGN source
density ($\approx$~7200~deg$^{-2}$, \citealt{bauer04}) than found in
optical surveys; see Fig.~1a for the sensitivity of the CDF-N to AGN
objects.

However, although deep X-ray surveys have proved stunningly effective
at identifying both obscured and unobscured AGN activity (see
\citealt{brandt05} for an overall review), there is clear evidence
that a large fraction of the AGN population remains undetected in the
\hbox{$<10$~keV} (observed frame) band probed by these surveys: (1)
about half of the X-ray background is unresolved at $>6$~keV
(\citealt{wors05}), (2) the obscured:unobscured AGN ratio is much
lower than that found for comparably luminous AGNs in the local
Universe (e.g.,\ \citealt{treist05}), and (3) few Compton-thick AGNs
($N_{\rm H}>10^{24}$~cm$^{-2}$) have been identified, even though they
comprise $\approx$~50\% of the AGN population in the local Universe
(e.g.,\ \citealt{tozzi06}; \citealt{page06}).

Many of these X-ray undetected AGNs are likely to be intrinsically
luminous sources that are heavily obscured ($N_{\rm
H}>3$--10~$\times10^{23}$~cm$^{-2}$) by an optically and geometrically
thick torus of gas and dust (e.g.,\ \citealt{wors05}; see Fig.~1a for
the detectability of the intrinsically luminous AGN in NGC~1068 in the
\hbox{CDF-N} observations). Although the primary emission from these
obscured AGNs will be absorbed by the torus and rendered weak or
invisible at X-ray--optical wavelengths, it will heat the dust grains
and re-emit this emission relatively isotropically in the infrared
(IR) waveband (rest-frame $>2$~$\mu$m; e.g.,\ \citealt{gran94};
\citealt{efstat95}). The intrinsically luminous but X-ray undetected
AGNs should therefore be detected in deep IR surveys; see
Fig.~1b.\footnote{High-resolution radio observations also provide
effective obscuration-independent selection of AGNs. However, most
radio surveys are either confined to relatively rare AGN source
populations, such as high-$z$ radio-loud AGNs or to AGN activity in
nearby galaxies (but see \citealt{donley05}). Future radio
observatories such as the Square Kilometer Array will provide an
extremely sensitive AGN census but will not be {\it fully} operational
for $>10$ years (e.g.,\ \citealt{jarvis04}); see
http://www.skatelescope.org/.} Since star formation can also produce
luminous IR emission it is necessary to distinguish between AGNs and
starburst galaxies. A number of recent studies have demonstrated that
IR colours can often distinguish between AGN and star-formation
activity using crude IR data, with the former having ``hotter'' IR
spectral energy distributions (SEDs; e.g.,\ \citealt{lacy04};
\citealt{stern05}). However, since a large fraction of the AGN
population in the local Universe has IR colors similar to those of
starburst galaxies, other indicators (e.g.,\ the detection of
high-excitation emission lines, hard X-ray emission from stacking
analyses of samples of objects, and luminous unresolved radio
emission) are also required to identify AGN activity from star
formation.

The primary aim of this article is to investigate the prospects for
selecting X-ray undetected AGN with {\it Spitzer} observations in the
ultra-deep CDF-N/GOODS-N field to increase the AGN census beyond that
identified from the X-ray data alone. Here we select X-ray undetected
AGN candidates using the {\it Spitzer} GOODS-N observations and then
search for the presence of hard X-ray emission using X-ray stacking
techniques. From these results we place constraints on the
demographics and properties of AGN activity in the IR galaxy
population.

\section{Searching for a large X-ray-undetected AGN population in the CDF-N/GOODS-N field}

The {\it Spitzer} GOODS observations (P.I.~M.~Dickinson) were
performed as part of the {\it Spitzer} Legacy program and cover
$\approx$~160~arcmin$^2$ of the most sensitive regions in each of the
{\it Chandra} deep fields (the CDF-N and the CDF-S). The IRAC
3.6--8~$\mu$m and MIPS 24~$\mu$m {\it Spitzer} observations are the
deepest across the sky, achieving typical 1~$\sigma$ sensitivities of
$\approx$~0.03--0.3~$\mu$Jy (3.6--8~$\mu$m) and $\approx$~5~$\mu$Jy
(24~$\mu$m). Here we focus on the CDF-N/GOODS-N field, which has the
deepest X-ray coverage to date. To provide maximum sensitivity to
X-ray faint AGN activity we restrict our analyses to the region within
5.5~arcmin of the {\it Chandra} aimpoint (95~arcmin$^2$), where the
X-ray sensitivity and angular resolution is optimal; see Figs.~18 \&
19 of \cite{alex03}. There are 223 X-ray sources, $\approx$~1800 MIPS
24~$\mu$m sources, and $\approx$~3400 IRAC 8~$\mu$m sources within
this region.

\begin{figure}[!t]
\plotfiddle{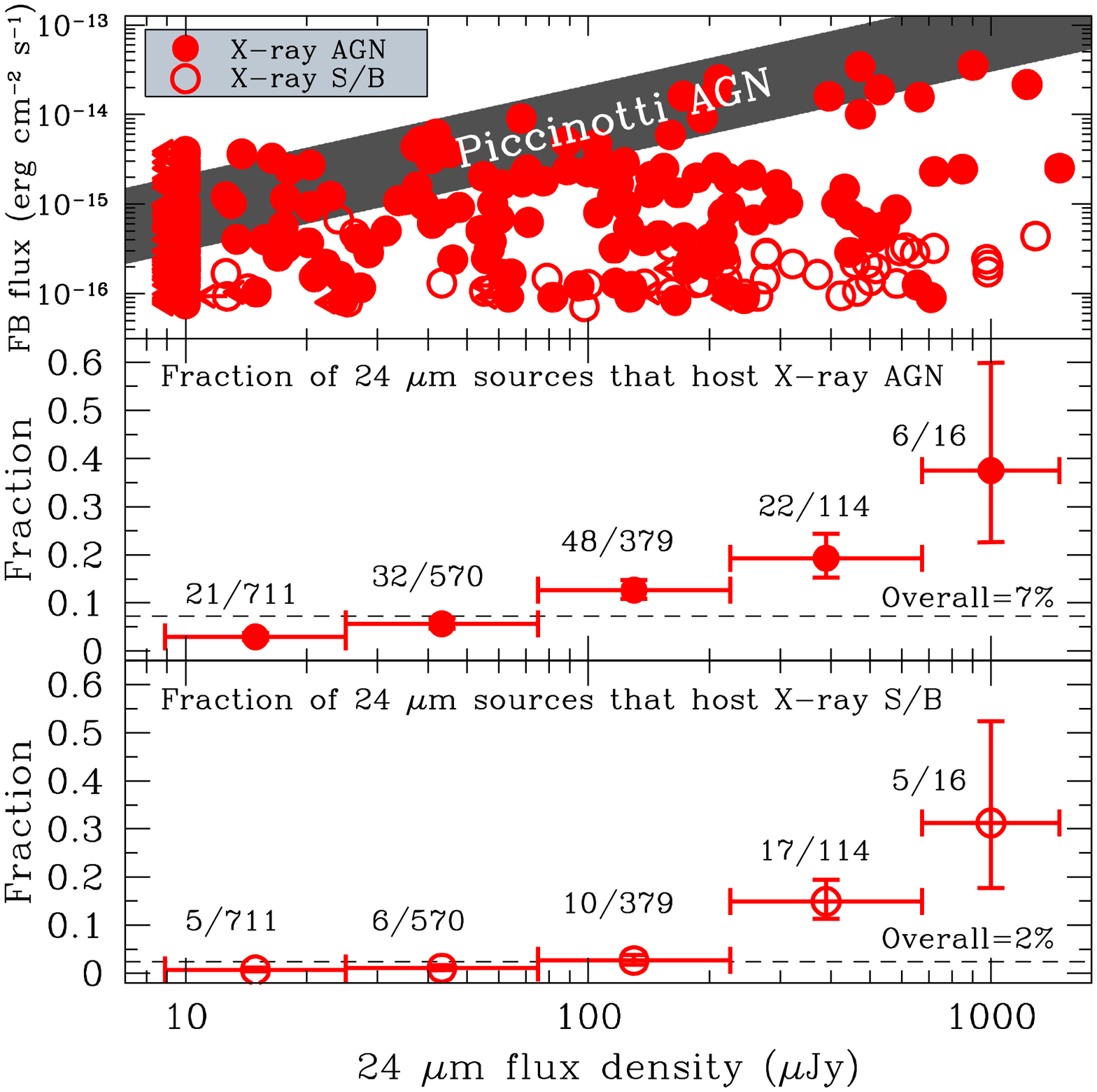}{3.8in}{0}{40}{40}{-110}{0}
\vspace{-0.8in}
\caption{24~$\mu$m flux density vs 0.5--8.0~keV flux (top), X-ray
  detected AGN fraction (middle), and X-ray detected starburst
  fraction (bottom). The shaded region shows the flux ratios of X-ray
  bright AGN (\citealt{picc82}) extrapolated to fainter fluxes. The
  decrease in the X-ray detected AGN fraction towards fainter
  24~$\mu$m flux densities is at least partially due to a relative
  decrease in the X-ray--IR flux ratio (i.e.,\ there is less area
  below the shaded region at faint 24~$\mu$m flux densities) and
  suggests that the majority of the X-ray undetected AGN population
  will have faint IR fluxes.}
\end{figure}

\subsection{X-ray Detected Infrared Galaxies}

The X-ray sources were matched to the IR sources on the basis of the
X-ray source positional uncertainties given in \cite{alex03}. The
positional uncertainites were linearly increased to give a 3\%
spurious source match fraction; extensive Monte-Carlo tests were used
to determine the spurious match fraction. Overall, 78\% and 87\% of
the X-ray sources were matched to 24~$\mu$m and 8~$\mu$m sources,
respectively. Using the multi-wavelength source classification of
\cite{bauer04} we distinguished between X-ray detected AGNs and X-ray
detected starbursts. The X-ray sources comprise 9.4\% of the 24~$\mu$m
source population (7.0\% X-ray AGNs and 2.4\% X-ray starbursts) and
5.6\% of the 8~$\mu$m source population (4.3\% X-ray AGNs and 1.3\%
X-ray starbursts).

Fig.~2 shows the X-ray vs 24~$\mu$m properties of the X-ray sources,
including the X-ray AGNs and X-ray starburst fraction for different
24~$\mu$m flux density bins. As expected, the majority of the X-ray
AGNs have X-ray--IR flux ratios higher than those of the X-ray
starbursts. Fig.~2 also shows a clear drop in the X-ray detected
fraction of IR sources towards the faintest IR fluxes. This is at
least partially due to the linear decrease in the observed X-ray--IR
flux ratio with 24~$\mu$m flux density, suggesting that the majority
of the X-ray undetected AGN population will have faint IR
fluxes.\footnote{This is clearly a simplistic prediction since Fig.~2
doesn't take into account possible changes in redshift and SED for
sources in the different 24~$\mu$m flux-density bins but it is at
least qualitatively similar to the conclusions drawn from the AGN
population synthesis model of \cite{treist06}.}

\subsection{Searching for X-ray Undetected AGNs}

Evidence for an X-ray undetected AGN population can be explored using
\hbox{X-ray} stacking techniques. The principle behind this approach
is to stack the X-ray data of individually undetected {\it Spitzer}
sources and determine the probability that the stacked X-ray signal
could be produced by the background (e.g.,\ \citealt{brandt01}; the
background is determined from a Monte-Carlo analysis of
source-excluded regions around the stacked sources. The expected
signature of a significant obscured AGN population is a detection in
the hard (2--8~keV) band and a comparatively flat X-ray spectral slope
($\Gamma<1.4$). By comparison, unobscured/low-luminosity AGN and
typical starburst galaxies have softer X-ray spectral slopes
($\Gamma\approx$~2; e.g.,\ \citealt{ptak99}; \citealt{george00}). We
note that neutron-star X-ray binary populations can also have flat
X-ray spectral slopes (e.g.,\ \citealt{colber04}), however, it seems
unlikely that the integrated X-ray emission from a starburst galaxy
will be dominated by this comparatively rare galactic population.

For the stacking analyses we have used the procedure outlined in
\cite{wors06} and stacked {\it Spitzer} galaxies in the soft
(0.5--2~keV), hard (2--8~keV), and full (0.5--8~keV) bands. The
background was determined using 1,000 Monte-Carlo trials for each
stacked source and the X-ray counts were measured using apertures
based on the 90\% encircled-energy radius. Stacking all of the X-ray
undetected IR sources we obtain highly significant detections
($>$~10~$\sigma$) in all three X-ray bands, corresponding to an
overall X-ray spectral slope of $\Gamma=1.4$ and suggesting that some
of the stacked sources are likely to be obscured AGNs (contrast with
the $\Gamma=2.0$ result obtained by \citealt{alex02} using shallower
15~$\mu$m data over a smaller area); this is further confirmed by a
3.5~$\sigma$ detection in the ultra-hard (4--8~keV) band. We also
stacked the data in the 6--8~keV band and obtained a non detection in
this band; the 3~$\sigma$ upper limit corresponds to $<$~20\% of the
6--8~keV background. If a substantial fraction of the X-ray background
is unresolved in the 6--8~keV band by current X-ray surveys, as
suggested by the analyses of \cite{wors05}, then the other
contributing sources must be either IR-faint ($f_{24\mu
m}<25$~$\mu$Jy) or comparatively rare in $\approx$~100~arcmin$^2$
regions.

A basic interpretation of the data in Fig.~2 is that X-ray undetected
AGNs will be predominantly detected at the faintest IR fluxes. To test
this simple prediction we stacked the X-ray data for individually
undetected sources in different 24~$\mu$m flux-density bins; see
Fig.~3a. The 2--8~keV stacked flux generally increases towards fainter
24~$\mu$m flux densities, suggesting an increase in the contribution
from obscured AGN activity, with the $f_{24\mu m}=$~75--225~$\mu$Jy
bin producing the strongest and hardest X-ray signal. However, the
softer X-ray spectral slope for sources with $f_{24\mu
m}=$~25--75~$\mu$Jy implies that starburst galaxies probably dominate
the stacked X-ray emission at the faintest IR fluxes. Furthermore, the
comparatively weak 2--8~keV flux suggests that there may be fewer
obscured AGNs than at $f_{24\mu m}=$~75--225~$\mu$Jy, possibly
indicating a genuine drop in the number of obscured AGNs at low IR
luminosities or at high redshifts ($f_{24\mu m}>$~75~$\mu$Jy
corresponds to $L_{\rm FIR}>10^{11}$~$L_{\odot}$ at $z=1$). These
results are qualitatively similar to those expected from the AGN
population synthesis model of \cite{treist06} and the interpretation
of shallower {\it Spitzer} and X-ray data by \cite{brand06}. More
detailed analyses are required to provide definitive conclusions.

\begin{figure}[!t]
\plottwo{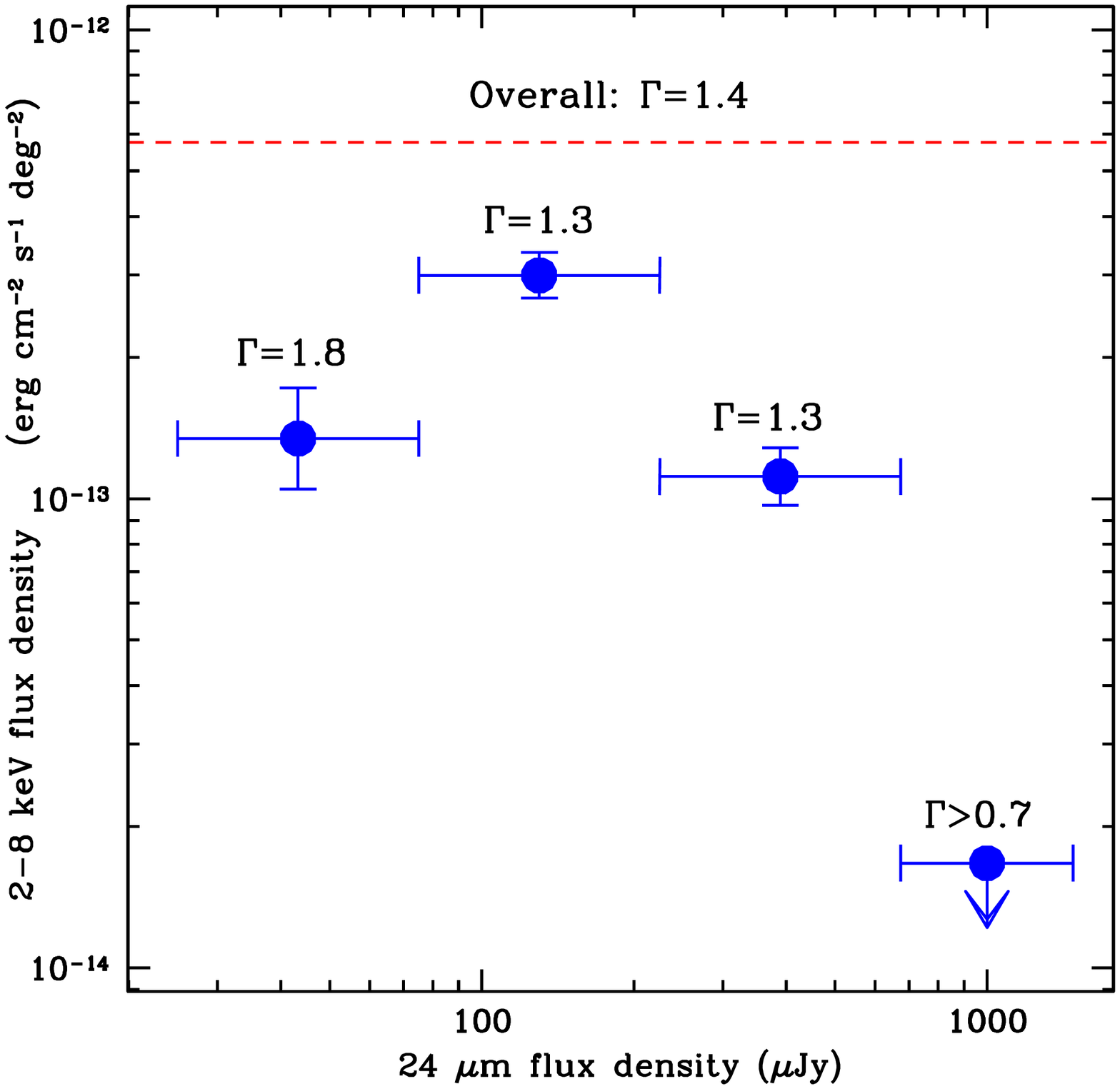}{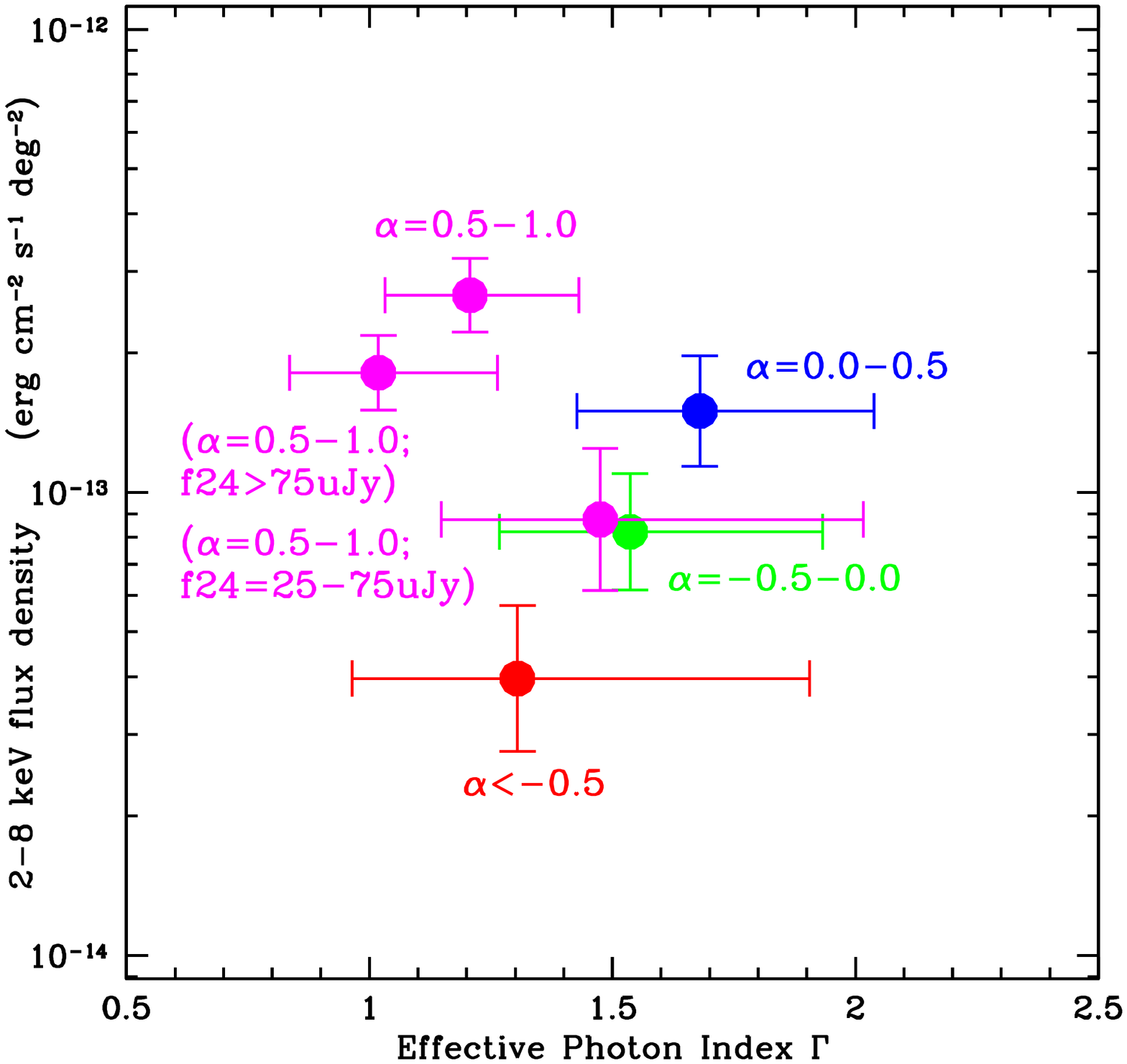}
\caption{(a) stacked 2--8~keV flux density vs 24~$\mu$m flux density
  for individually undetected {\it Spitzer} galaxies; the X-ray
  spectral slope for each 24~$\mu$m flux-density bin is also
  shown. The strongest and hardest X-ray signal is produced by $f_{\rm
  24\mu m}=$~75--225~$\mu$Jy sources. The dashed line indicates the
  overall stacked X-ray signal obtained by stacking all of the {\it
  Spitzer} galaxies. (b) stacked 2--8~keV flux density vs X-ray
  spectral slope for individually undetected {\it Spitzer} galaxies
  with different IR spectral slopes ($\alpha$). The results for the
  $\alpha=$~0.5--1.0 sources are also given for different flux ranges
  ($f_{24\mu m}=$~25--75~$\mu$Jy and $f_{24\mu m}>75$~$\mu$Jy); the
  strongest and hardest X-ray signal comes from the $f_{24\mu
  m}>75$~$\mu$Jy sources.}
\end{figure}

We can also explore the average X-ray properties of the {\it Spitzer}
sources as a function of IR spectral slope ($\alpha$, where
$f_{\nu}$~$\alpha$~$\nu^{\alpha}$). In the local Universe many AGNs
have flat IR spectral slopes while starburst galaxies are often
distinguished by steeper IR spectral slopes. \cite{alonso06}
investigated the detailed properties of {\it Spitzer} galaxies with
flat IR spectral slopes ($\alpha<-0.5$) in the 1~Ms CDF-S field and
deduced that only $\approx$~50\% of these IR-selected AGNs are
detected in the X-ray band. Although the flat IR spectral slopes of
these sources clearly indicate that they host AGN activity, they are
comparatively rare ($\approx$~800 sources deg$^{-2}$, including both
X-ray detected and X-ray detected AGNs) and unlikely to dominate the
unresolved $>8$~keV background. As a comparison, the range of IR
spectral slopes for the X-ray detected AGN in the CDF-N/GOODS-N
dataset is $\alpha=-2.0$--1.0, and only $\approx$~30\% have
$\alpha<-0.5$, indicating that many have IR spectral properties more
consistent with those of starburst galaxies. Since the aim of the
study here is to provide the most complete census of AGN activity, we
have stacked the X-ray data of {\it Spitzer} galaxies over a wide
range of IR spectral slopes ($\alpha<-0.5$, $\alpha=-0.5$--0.0,
$\alpha=$~0.0--0.5, $\alpha=$~0.5--1.0). The X-ray spectral slope and
2--8~keV flux density of the stacked sources are shown in Fig.~3b. All
are significantly detected in the 2--8~keV band but the strongest and
hardest X-ray signal comes from the sources with the steepest IR
spectral slope ($\alpha=$~0.5--1.0). The steep IR spectral slope of
these sources does not obviously imply the presence of AGN activity
but the hard X-ray signal ($\Gamma=$~1.2) indicates that a
considerable fraction of the sources host obscured AGN
activity. Motivated by the discovery that sources with $f_{24\mu
m}>$~75~$\mu$Jy show greater evidence for X-ray undetected AGN
activity (see Fig.~3a), we further stacked the sources in two
different 24~$\mu$m flux bins ($f_{24\mu m}=$~25--75~$\mu$Jy and
$f_{24\mu m}>$~75~$\mu$Jy). The flatter X-ray spectral slope
($\Gamma=1$) and larger 2--8~keV flux density of the $f_{24\mu
m}>$~75~$\mu$Jy sources with $\alpha=$~0.5--1.0 clearly indicates that
the majority of the obscured AGNs reside in this population.

\section{Discussion}

The X-ray stacking analyses provide evidence that a large X-ray
undetected AGN population does indeed reside in the {\it Spitzer}
galaxy population. The dominant AGN population has an IR spectral
slope of $\alpha=$~0.5--1.0 and $f_{24\mu m}>$~75~$\mu$Jy. Their IR
colors are more similar to starburst galaxies than AGNs but X-ray
stacking analyses indicate that they host obscured AGN activity. We
can gain further insight into the properties of these sources by
investigating their median redshifts, luminosities, and source
densities.

The median spectroscopic redshift of the $\alpha=$~0.5--1.0 {\it
Spitzer} galaxies is $z\approx$~0.8, which corresponds to an median
luminosity of $L_{\rm FIR}\approx10^{11}$~$L_{\odot}$; the majority of
the redshifts were taken from \cite{wirth04}. Assuming this
median redshift, the 2--8~keV luminosity of the stacked sources with
$\alpha=$~0.5--1.0 and $f_{24\mu m}>$~75~$\mu$Jy is
$\approx10^{41}$~erg~s$^{-1}$ (rest-frame 3.6--14.4~keV). This X-ray
luminosity should be considered a lower limit since (1) many of the
stacked galaxies will not be AGNs and their inclusion in the stacking
analysis will decrease the average 2--8~keV flux, and (2) the
luminosity has not been corrected for the presence of instrinsic
absorption. These two issues are briefly considered below.

We cannot directly determine the ``dilution'' to the hard-band X-ray
signal from non-AGN activity, however, we can place constraints from
the non-detection of individual sources in the X-ray data. For
example, if the X-ray undetected AGN represented just 1\% of the
stacked IR galaxies but contributed all of the 2--8~keV flux then they
would have individual 2--8~keV luminosities $\approx$~2 orders of
magnitude higher and would be individually detected in the X-ray
data. Therefore, since the stacked 2--8~keV flux is an order of
magnitude below the individual X-ray sensitivity limit, these X-ray
undetected AGNs are likely to represent $>10$\% of the stacked IR
galaxies. This indicates an average {\it observed} (i.e.,\ uncorrected
for absorption) 2--8~keV luminosity of
$\approx10^{41}$--$10^{42}$~erg~s$^{-1}$ and places lower limits on
the X-ray undetected obscured AGN source density of
$\approx$~500--5,000~deg$^{-2}$ (for AGNs with $\alpha=$~0.5--1.0 and
$f_{24\mu m}>$~75~$\mu$Jy).

A crude constraint on the amount of absorption towards the AGNs can be
determined from the average X-ray spectral slope for a given AGN model
(e.g.,\ an observed X-ray spectral slope of $\Gamma=$~1.0 from a
$z\approx$~0.8 AGN would imply $N_{\rm
H}\approx$~$3\times10^{22}$~cm~s$^{-2}$ for a simple intrinsic AGN
power-law spectrum of $\Gamma=2.0$). Since the soft-band flux almost
certainly has a large contribution from star-formation activity, this
approach effectively provides a lower limit on the true absorbing
column density. Another approach, albeit very uncertain, is to
estimate the {\it intrinsic} (i.e.,\ corrected for absorption) X-ray
luminosity of the AGNs from the IR luminosities and then compare this
to the observed X-ray luminosity. Under the assumption that 10--100\%
of the IR luminosity is powered by an AGN, the median X-ray luminosity
of these sources estimated from the \cite{elvis94} SED is
$\approx10^{42}$--$10^{43}$~~erg~s$^{-1}$, suggesting large amounts of
absorption ($N_{\rm H}\gg10^{23}$~cm$^{-2}$).

Although these analyses on the properties of the X-ray undetected AGNs
are crude, their estimated properties are qualitatively consistent
with those required by \cite{wors05} to produce the unresolved X-ray
background (i.e.,\ $z\approx$~0.5--1.5, $N_{\rm
H}\approx10^{23}$--$10^{24}$~cm$^{-2}$, $L_{\rm
X}\approx10^{42}$--$10^{43}$~erg~s$^{-1}$). Interestingly, the median
redshift, IR luminosity, and X-ray luminosity range of the X-ray
detected AGNs with $\alpha=$~0.5--1.0 are also similar to those
estimated for the X-ray undetected AGNs with $\alpha=$~0.5--1.0. This
is entirely consistent with that expected for the unified AGN model,
where the X-ray undetected AGNs represent the more heavily obscured
counterparts of the X-ray detected AGNs. Detailed observations of
individual objects are now required to elucidate the properties of
this X-ray undetected AGN population.


\acknowledgements I thank the Royal Society for support and my
collaborators on this project so far: R.~Chary, M.~Dickinson,
F.~Bauer, W.~Brandt, A.~Fabian, M.~Worsley, and the {\it
Spitzer}-GOODS team.


\end{document}